\documentclass[%
reprint,
superscriptaddress,
amsmath,amssymb,
aps,prl
]{revtex4-1}

\usepackage{graphicx}% Include figure files
\usepackage{dcolumn}% Align table columns on decimal point
\usepackage{bm}% bold math
\usepackage{hyperref}% add hypertext capabilities
\usepackage{float}
\usepackage{cases}
\usepackage{cleveref}
\usepackage{wasysym}
\usepackage[makeroom]{cancel}
\usepackage{scalerel}
\usepackage{lipsum}
\hypersetup{
	colorlinks,
	linkcolor={red},
	citecolor={blue},
	urlcolor={black}
}

\begin{document}

	\preprint{APS/123-QED}
	\sloppy
	\allowdisplaybreaks
% 	\title{Thermodynamic Valence Control in the Self-Assembly of DNA-Coated Droplets}
	\title{DNA self-organization controls valence in programmable colloid design}
	%Equilibrium Colloidal Valence as an Emergent Property of Molecular Adhesion}
	% Mobile flexible linkers spontaneously self-organize to prescribe particle valence

	% Force line breaks with \\
	% \thanks{S.A. and F.K.C. contributed equally to this work.}%

	\author{Angus McMullen}
	\affiliation{Center for Soft Matter Research, New York University, New York, NY, USA}
	%\thanks{S.A. and F.K.C. contributed equally to this work.}
	\author{Sascha Hilgenfeldt}
	\affiliation{Mechanical Science and Engineering, University of Illinois, 
Urbana-Champaign, Illinois 61801, USA}
	%\thanks{S.A. and F.K.C. contributed equally to this work.}
%	\affiliation{Mechanical Science and Engineering, University of Illinois, 
%Urbana-Champaign, Illinois 61801, USA}

	\author{Jasna Brujic}
	\affiliation{Center for Soft Matter Research, New York University, New York, NY, USA}
	\email{jasna.brujic.nyu.edu}
	%
	%  \altaffiliation[Also at ]{Physics Department, XYZ University.}%Lines break automatically or can be forced with \\
	%  \email{Second.Author@institution.edu}
	% \affiliation{%
	%  Authors' institution and/or address\\
	%  This line break forced with \textbackslash\textbackslash
	% }%

	% \collaboration{MUSO Collaboration}%\noaffiliation

	%  \homepage{http://www.Second.institution.edu/~Charlie.Author}
	% \affiliation{
	%  Second institution and/or address\\
	%  This line break forced% with \\
	% }%
	% \affiliation{
	%  Third institution, the second for Charlie Author
	% \affiliation{%
	%  Authors' institution and/or address\\
	%  This line break forced with \textbackslash\textbackslash
	% }%

	% \collaboration{CLEO Collaboration}%\noaffiliation

	\date{\today}% It is always \today, today,

\begin{abstract}
Just like atoms combine into molecules, colloids can self-organize into predetermined structures according to a set of design principles. 
Controlling valence---the number of inter-particle bonds---is a prerequisite for the assembly of complex architectures.
The assembly can be directed via solid `patchy' particles with prescribed geometries to make, for example, a colloidal diamond. 
We demonstrate here that the nanoscale ordering of individual molecular linkers can combine to program the structure of microscopic assemblies. 
Specifically, we experimentally show that covering initially isotropic microdroplets with $N$ mobile DNA linkers results in spontaneous and reversible self-organization of the DNA into $Z(N)$ binding patches, selecting a predictable valence. 
We understand this valence thermodynamically, deriving a free energy functional for droplet-droplet adhesion that accurately predicts the equilibrium size of and molecular organization within patches, as well as the observed valence transitions with $N$. 
Thus, microscopic self-organization can be programmed by choosing the molecular properties and concentration of binders.
These results are widely applicable to the assembly of any particle with mobile linkers, such as functionalized liposomes or protein interactions in cell-cell adhesion.

	\end{abstract}

	\maketitle

	%\tableofcontents

Building blocks encoded with assembly rules harness thermal energy to put themselves together in a process called self-assembly \cite{Whitesides2002, rogers2016using}. 
These elements can be proteins \cite{pyles2019controlling, lai2014structure}, DNA \cite{rothemund2006folding, douglas2009self, winfree1998design, nykypanchuk2008dna}, or colloids \cite{rogers2015programming, he2020colloidal, nykypanchuk2008dna, casey2012driving, lin2017clathrate}. 
Akin to atoms and molecules, colloidal particles with well-defined shapes and interactions self-organize into bulk crystalline phases that minimize the free energy \cite{manoharan2015colloidal, wang2017colloidal, oh2019colloidal, ducrot2017colloidal, varilly2012general, angioletti2012re, grunwald2014patterns}.
More complex objects with non-repeating structures, such as protein folds or aperiodic crystals, require a prescribed limit to particle valence \cite{ VanAnders2014, lu2015superlattices}. 
A fundamental goal is to fabricate structures with important technological applications \cite{cademartiri2015programmable}. 
For example, colloidal self-assembly into a diamond lattice \cite{he2020colloidal} or a quasicrystal \cite{haji2009disordered, engel2015computational} is expected to exhibit photonic band gaps due to the materials' interaction with light \cite{Yablonovitch1989, Yablonovitch1991}. 
At its most complex, self-assembly of biological cells is a crucial part of the development of a living organism  \cite{blanchard2009tissue}.

Experimentally, valence control can be achieved by designing anisotropic sticky particles with patches to create colloidal clusters \cite{Wang2012, chen2012janus, chen2011supracolloidal} or DNA origami that specifies the bond orientation \cite{Zhang2018,Zion2017}. 
Mixing particles with a given size and number ratio can result in steric valence control \cite{perry2015two}. 
Other proposed methods include the self-organization of nematic shells on spheres \cite{Kralj2011,Nelson2002} or the arrested phase separation of lipids on droplet surfaces \cite{pontani2013immiscible}. 
These processes are complex to experimentally realize, feature slow assembly kinetics due to the necessity of patch-to-patch binding, and require extensive purification \cite{Wang2012}.  

Unlike solid particles, droplets \cite{Feng2013, McMullen2018, pontani2016cis, dlamini2021self}, lipid vesicles \cite{Bachmann2016,And2007,Chung2013,Parolini2016,Parolini2015,Shimobayashi2015}, and biological cells \cite{collinet2021programmed, hilgenfeldt2008physical, bell1984cell, bell1978models} allow any sticky binders to freely diffuse at the interface and segregate into adhesions with their neighbors. 
If the particles are Brownian or mobile, they can rearrange even after binding to reach the most favorable valence and geometry, avoiding kinetic bottlenecks. Angioletti-Uberti \textit{et al.} theoretically proposed that mobile ligands coupled with an additional repulsive potential---such as a steric brush---could yield colloidal valence selection in the bulk \cite{Angioletti-Uberti2014}. 
More generally, the mobility and reversibility of linker binding between particles allows the system to optimize its equilibrium structure according to the laws of statistical mechanics. 
Not only is this strategy more robust than directed irreversible assembly, but it enables colloidal design based on the properties of molecular binders.  

Here, we derive and experimentally validate the free energy functional for droplet-droplet adhesion and predict the consequent thermodynamically stable valence for given control parameters. 
Moreover, we show that droplets recover their equilibrium valence in a matter of minutes after their bonds are broken. Our results are applicable to any functionalized particles with mobile binders, showing that molecular properties and concentration are sufficient to predetermine valence. 
Emulsions serve as a template for programmable solid materials because the droplets can be readily polymerized at any stage of the self-assembly process \cite{middleton2019optimizing, oh2020photo}. 

\begin{figure*}[t!]
%   \centering
   \includegraphics[width=\textwidth]{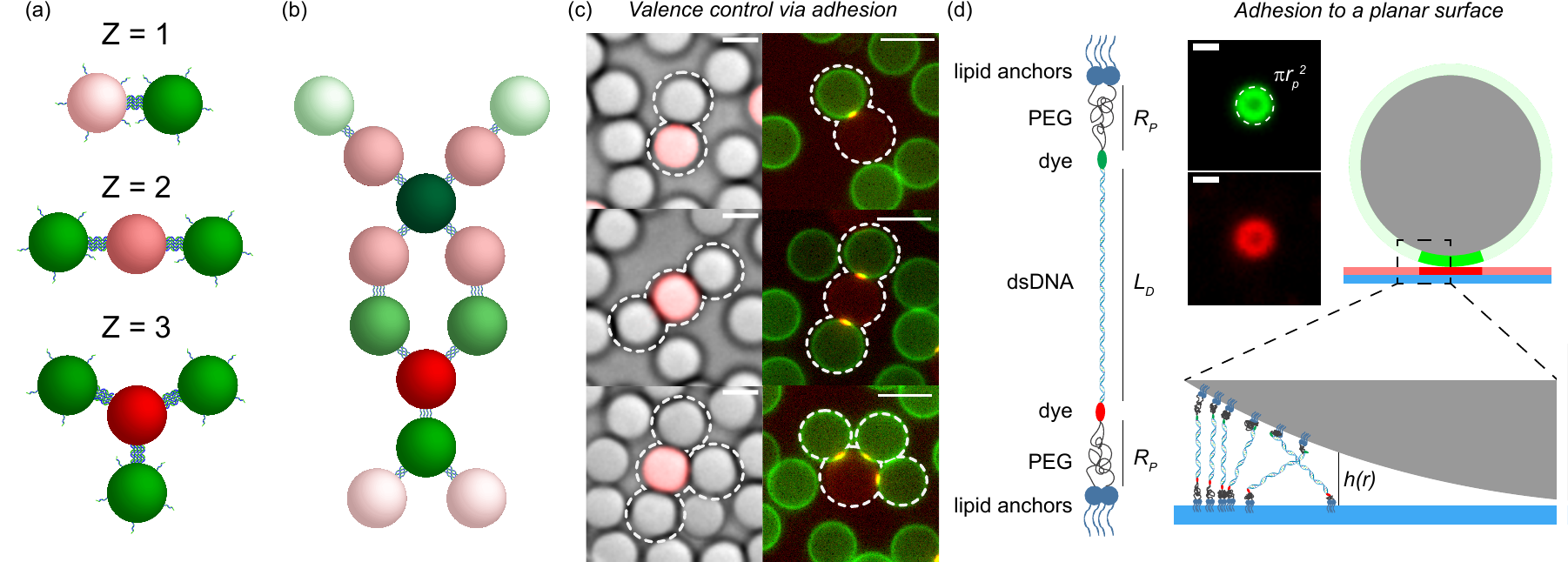} % requires the graphicx package
   \caption{Droplet valence via complementary DNA linkers. (a) A schematic of droplet valence $Z=1,2,3$ shows their assembly into complex structures, e.g. (b). (c) Experimental data shows brightfield images of $R_0$ = 2 $\mu$m droplets with $Z=1,2,3$, while equivalent fluorescence images reveal dense DNA adhesion patches (yellow). Scale bars are 5\,$\mu$m. (d) A schematic of a single bound DNA complex, as well as their recruitment into the adhesion of a droplet with a solid surface decorated with mobile linkers. Image inset shows the fluorescence colocalization of complementary DNA on the droplet (green) and the surface (red) inside the adhesion patch area $\pi r_p^2$. Scale bar is 1\,$\mu$m.  
}
   \label{fig:intro}
\end{figure*}

\section*{Results}

We consider pairwise droplet binding through semi-flexible DNA linkers. 
Droplets are decorated with a double stranded DNA (dsDNA) tether to a 20-base single stranded sticky end (species $A$, red in Fig\,\ref{fig:intro}) or its complement (species $A^\prime$, green in Fig\,\ref{fig:intro}). 
Each resulting molecular bond consists of the DNA and two PEG coils attached to the droplets via lipid anchors, see Fig\,\ref{fig:intro}d. 
Mixing both species at a given number of DNA molecules per droplet, $N$ and $N^\prime$, the droplets form a well-defined number of binding patches, i.e. valence $Z$, which remains fixed despite frequent collisions with neighboring droplets. 
Mobile DNA molecules are recruited into localized adhesion patches, shown in bright fluorescent yellow in Fig\,\ref{fig:intro}c. 
As the surface DNA density increases (here we increase $N$ at fixed $N^\prime$), so does droplet valence. 

To determine whether this valence control is kinetic or thermodynamic in origin, we aim to derive the free energy of individual patch formation. 
Therefore, we experimentally measure the equilibrium patch size and DNA density profiles as a function of DNA coverage, linker length, and droplet size. 
For precise visualization, these patches are formed between an $A^\prime$ droplet (radius $R_0=2.9\mu$m, green) and a complementary $A$-labeled hydrophobic glass surface (red), as shown in the fluorescence images of Fig\,\ref{fig:intro}d. 
The co-localization of fluorescence inside the patch indicates that the surface bound DNA is able to freely diffuse laterally.

\begin{figure*}[t]
  \centering
  \includegraphics[width=\textwidth]{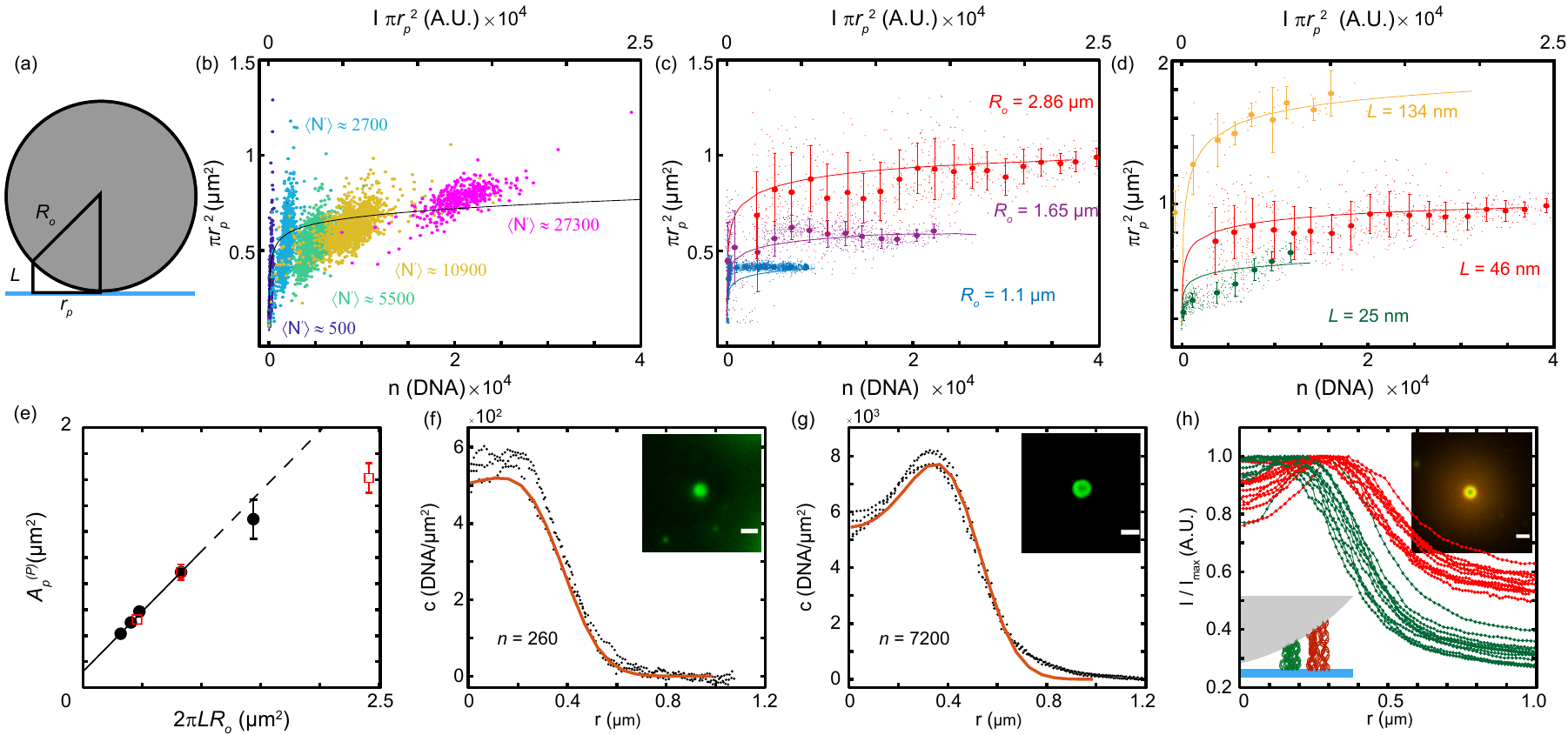} % requires the graphicx package
  \caption{%Free energy of 
  Droplet-surface adhesion.
  %[[sugest this because this is about more than just free energy]] 
  (a) A schematic of the patch geometry between a droplet and a surface. (b)  Patch area $\pi r_p^2$ grows as a function of intensity $I \pi r_p^2$ and corresponding $n ({\rm DNA})$ for $R_0=2.1 \mu$m droplets in agreement with the theory (solid line) with no adjustable parameters. Increasing bulk DNA concentration increases droplet coverage $N$ with a spread shown by the colors. 
  The model captures the increase in average area with droplet radius $R_0$ in (c) and DNA complex length $L$ in (d). (e) The maximum experimental area grows as $\approx 2\pi L  R_0$ when varying $R_o$ and fixing $L = 46$\,nm (black circles) or when varying $L$ with fixed $R_0=2.86 \mu$m (red squares), confirming the spherical geometry in (a). The nonzero intercept %(0.13 $\mu^2$)
  corresponds to the PSF blur. 
  %[[took out the number because it is not quanitatively the PSF area]] 
  (f,g) For each patch size, theory also predicts the DNA density profile $c(r)$ (red line), in excellent agreement with experimental profiles (black dots) for low (f) and high (g) density patches, shown in the insets. (h) Normalized intensity profiles from droplets decorated with two binder lengths, 25\,nm (green) and 46\,nm (red), reveal concentric rings (inset), as predicted by geometry. Scale bars are 1\,$\mu$m. }
  \label{fig:AvsI}
\end{figure*}

We vary the amount of DNA on the droplets, $\langle N^\prime\rangle$, and measure the resulting circular patch area, $A_p=\pi r_p^2$ as well as the integrated intensity of the patch $I$ (see Figure \,\ref{fig:AvsI}b). 
The number of DNA molecules $n\propto IA_p$ recruited into the patch is then calculated for every patch. 
Plotting $A_p$ versus $n$ shows a fast patch growth at small $n$ asymptoting to a plateau value $A_p^{(P)}$ as DNA molecules pack into the patch.

Next, we show that this limit in patch size arises because the droplets do not deform to accommodate flat adhesion patches, but remain spherical.
The binder length is limited to $L=L_D+2R_P$, where $L_D$ is the contour length of the DNA and $R_P$ is the equilibrium length of the PEG coils, which remains nearly unextended \cite{mandelkern1981dimensions} (see Methods for details). 
Thus, the maximum patch area is given by simple geometry as $A_p^{(P)}\approx 2\pi L R_0$ to excellent approximation (as $L\ll r_p \ll R_0$), see Fig.\,\ref{fig:AvsI}a. 

Indeed, in Figure\,\ref{fig:AvsI}c,d we show that the binned average plateau value from experiment increases when varying either $R_0$ or $L$, and Fig.\,\ref{fig:AvsI}e demonstrates that the increase is quantitatively explained by the assumption of undeformed droplet geometry. 
The predicted linear trend of  $A_p^{(P)} = 2\pi L R_0$  holds with a slope of $0.92$ instead of $1$. 
\textcolor{black}{The experiment using the largest-length binders deviates from this trend because even for the highest binder concentrations the DNA molecules at the edge of the patch are not stretched to their contour length.}
The largest DNA length deviates from the trend because the molecules can no longer be considered a rigid rods. 
The nonzero intercept is due to the point spread function (PSF) blur of the microscope ($\approx 300nm$). The linear scaling confirms that the droplet surface tension of 10\,mN/m is too large to make droplet deformation favorable at these concentrations of DNA. 
This result is contrary to the common assumption that binding energy competes with deformation energy~\cite{pontani2012biomimetic, Feng2013, Zhang2017}. The fact that droplets remain spherical implies that our results are equally valid for solid particles with fluid interfaces \cite{van2013solid, rinaldin2019colloid}.  

We derive the free energy of adhesion ${\cal F}_p$ to explain the patch growth with $n$ with no adjustable parameters, as shown in Fig.~\ref{fig:AvsI}b. 
For a given number $n$ of binders in a patch, we write ${\cal F}_p= {\cal F}_s+{\cal F}_{int}$, where ${\cal F}_s$ is the spring energy of the binders stretched or compressed to variable length and ${\cal F}_{int}$ is their interaction energy. 
These can be written as integrals over the circular patch area $A_p$
\begin{align}
{\cal F}_{s} &= \int_{A_p} c(r) s(r) (2\pi r) dr, \\ {\cal F}_{int} & = 
\int_{A_p}(c^2(r)/c_m) g(r) (2\pi r) dr\,, 
\end{align}
where $s(r)=s(h(r))$ and $g(r)=g(h(r))$ are the spring energy and interaction energy for single binders connecting two surfaces of distance $h$, respectively. 
For small circular patches, the undeformed droplet shape implies $h(r)=r^2/(2R_0)$, as noted above. 
The precise form of $s(r)$ and $g(r)$ depends on the molecular properties of the binder and we shall give them in units of $k_BT$ below. 
In our case, the binder is a series of a flexible nonlinear spring (the double-stranded DNA of length $L_D$) and two ideal-chain springs (PEG molecules of unextended size $R_P\approx 2.4$nm) (see Fig.~\ref{fig:intro}d).

For rod-like double-stranded DNA binder composites of maximum length $L$ and width $d_D$, we define a characteristic interaction concentration $c_m=1/(L d_D)$. 
A straightforward application of Onsager volume exclusion \cite{onsager1949effects} (see Methods) then yields an interaction energy
\begin{equation}
g(h)=2(L+d_D-h)/L\,.  
\label{gofz}
\end{equation}
The spring energy of a composite binder is 
\begin{equation}
s(h)= s_D(h) + 2s_P(h) \,,
\label{stot}
\end{equation}
where $s_P$ is the ideal spring energy of a PEG coil, while $s_D$ is taken from the theory of flexible FENE springs \cite{winkler2003deformation} (see Methods). 
The contributions $h_P$ and $h_D=h-h_P$ for the PEG and DNA, respectively, to the total surface-to-surface distance $h$ are obtained from the condition of equal forces in the PEG-DNA-PEG series of springs (see Methods).

\begin{figure*}
  \centering
  \includegraphics[width=\textwidth]{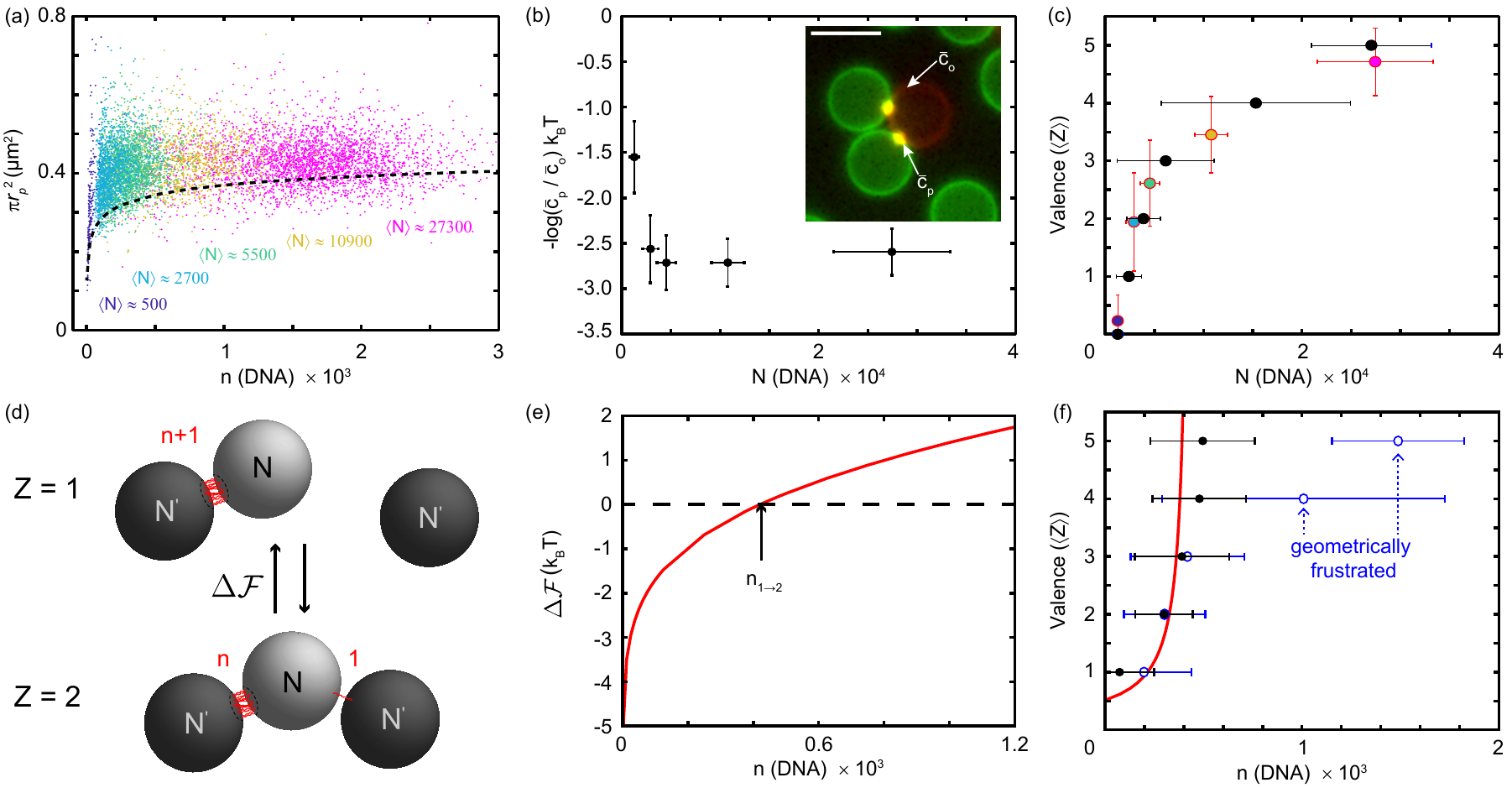} % requires the graphicx package
  \caption{Valence control in droplet self-assembly. (a) Patch size quickly reaches the geometric plateau with $n$ in droplet-droplet adhesion ($R_0 = 2.25 \mu$m). Increasing bulk DNA concentration creates denser patches of a similar size (see colors), in agreement with theory (dashed line). 
  (b) The ratio of concentrations outside and inside the patch, $-\log(\bar{c}_p / \bar{c}_o)$, reveals a free energy of adhesion per molecule on the order of $k_BT$, independent of droplet coverage $N$. (c) During self-assembly, this allows for patch rearrangements into an average droplet valence (symbol colors correspond to (a)) that increases with $N$. Conversely, a given valence is realized by a defined range of $N$ (black).
  (d) A schematic illustrating the valence transition from $Z_{1\to 2}$, which occurs in (e) when the free energy difference $\Delta {\cal F}$  between the two states becomes positive, above the patch population $n_{1\to 2}$. (f) Measured valence (blue, open) and measured valence after removal of droplets with supersaturated adhesion patches (black, closed) as a function of $n$. The latter is in excellent agreement with the theoretical prediction (red) for all $Z$.}
  \label{fig:valence}
\end{figure*}

Given $s(h(r))\equiv s(r)$ and $g(h(r))\equiv g(r)$, minimization of ${\cal F}_p$ with respect to the concentration profile (under the requirement 
of constant $n$) is straightforward and yields
\begin{equation}
c(r) = \frac{c_m}{2g(r)}\left(\frac{2n}{c_m \pi r_p^2 \Gamma_0} + \frac{\Gamma_1}{\Gamma_0} -s(r)\right) \,,
\label{cexplicit}
\end{equation}
where 
\begin{equation}
\Gamma_k \equiv \Gamma_k(r_p)= \frac{1}{\pi r_p^2} \int_0^{r_p} \frac{s^k(r)}{g(r)} (2\pi r)dr\,,\label{gammadef}
\end{equation}
with $k=0,1$. As $c(r_p)=0$, the concentration profile immediately determines $r_p$ through
 \begin{equation}
\frac{2n}{c_m\pi r_p^2} = s(r_p) \Gamma_0 - \Gamma_1\,. \label{rpeq}
\end{equation}

With all parameters known, we directly compare the equilibrium patch size as a function of $n$ with experimental data in Fig.~\ref{fig:AvsI}b-d and find excellent agreement as $R_0$ and $L_D$ are varied. 
Discrepancies for the smallest droplet size and shortest linker length are likely due to limited experimental resolution. 

Via Eq.~\eqref{cexplicit}, the theory also predicts non-trivial radial profiles in DNA density within the patch. 
In patches that are significantly smaller than $A_p^{(P)}$, experimental measurements (black circles in Fig.~\ref{fig:AvsI}f) show that DNA is distributed homogeneously and the intensity decay at the edge is dominated by the PSF of the microscope. 
As DNA crowds inside larger patches with $A_p \lesssim A_p^{(P)}$, it preferentially migrates towards the edge to form a ring of extended rods avoiding intermolecular interaction penalties. 
These profiles show a non-zero radius peak in DNA density, see Fig.~\ref{fig:AvsI}g. 
Both uniform and ring-shaped profiles are predicted from Eq.~(\ref{cexplicit}), after convolution with the PSF, to excellent accuracy with no adjustable parameters (red lines in Fig.~\ref{fig:AvsI}f,g). 
The disk to ring shape transition occurs when the DNA repulsion term dominates over the entropic loss of DNA extension. 
As $\langle N^\prime\rangle$ is further increased, the ring becomes denser and gradually fills up the patch.  

Further confirming the spherical droplet geometry, Figure~\ref{fig:AvsI}h shows that droplets decorated with DNA binders of two different lengths form adhesions organized into concentric rings, as expected. 
This result is analogous to the spontaneous formation and segregation of protein binder rings in the biological immune response \cite{chakraborty2010statistical, qi2001synaptic}, albeit via a different physical mechanism.

Next, we show how the free energy of adhesion explains valence transitions in droplet self-assembly. 
Figure~\ref{fig:valence}a shows that droplet-droplet adhesion patch size varies with $n$ similarly to droplet-surface adhesion. 
In these experiments, the $A$ species droplets were labeled with a range of DNA numbers (see legend), while the $A^\prime$ droplets were decorated at the highest DNA density. 
$A$ droplets were mixed with a large excess of the $A^\prime$ species to ensure that $A$ droplets reach their maximum valence. 

Under the same assumption of undeformed spherical droplets, the theory described above for droplet-substrate binding applies with the simple change of halving $L$ (see Methods). \textcolor{black}{All predictions for patch size and binder density profiles remain of the same form.
Figure~\ref{fig:valence}a shows that good agreement is again obtained with no adjustable parameters.} 
The slight underestimate of the patch size by the theory is likely due the difficulty of estimating patch size when the patch is oriented perpendicular to the imaging plane.
Note that experimental patches come from all droplet valences, while the $A_p(n)$ theory considers monovalent droplets.  

A prerequisite for thermodynamic valence control is the ability of DNA molecules to rearrange between patches and achieve their equilibrium configuration. 
As a measure of effective binding energy \cite{bell1978models}, we determine the logarithm of the ratio of average DNA concentrations inside and outside the patch, $-\log (\bar{c}_p / \bar{c}_o)$, shown in Figure\,\ref{fig:valence}b. 
This energy proves to be on the order of $k_BT$, largely independent of the DNA coverage $N$ on the droplet surfaces, revealing that a significant fraction of DNA remains unbound.  
Note that the energy value for the first data point may be unreliable since it represents droplet intensities close to the background intensity. 
This result indicates that, even though the nominal binding energy of our paired DNA sticky ends is $\sim 30 k_BT$ \cite{santalucia2004thermodynamics}, steric confinement in the patch \cite{jonchhe2020duplex} together with the effects of crowding and stretching between undeformed droplet interfaces lowers it sufficiently to allow for thermal equilibration between the interior and exterior of patches. 

Both $A$ and $A^\prime$ droplets are prepared with a small ($<R_0$) gravitational height, so that they can interact and bind in a single 2D layer against a microscope coverslip (see Methods). 
After self-assembly reaches a steady state, we measure the valence of each $A$ particle and quantify the $N$ of each droplet and the $n$ of each patch from the fluorescence signal.
We find that droplets select their valence according to the DNA coverage of their $A$ partners, as shown in Figure\,\ref{fig:valence}c. 
We either measure a fixed valence $Z$ and the corresponding average of the DNA per droplet (blue markers), or we measure the average DNA per droplet in a single sample and the corresponding average $Z$ (red outlined markers).  
The fact that both methods yield the same increase of valence with $N$  indicates that DNA coverage is sufficiently uniform between droplets for bulk valence selection. 
Valence increases linearly at first and then gradually saturates towards the crystalline packing limit $Z=6$ in the 2D layer of droplets. 

%The average DNA per patch as a function of valence, seen in Fig.\,\ref{fig:valence}(d) (blue markers), shows a similar trend. 

The high reproducibility of the valence results suggests that the states of varying valence are thermodynamic equilibrium states, or close to such. 
Due to the small effective Boltzmann factors shown in Fig.~\ref{fig:valence}b, molecules are not kinetically trapped and are readily exchanged between patches and their exterior.
\textcolor{black}{Note that the relation between valence and droplet coverage of Fig.~\ref{fig:valence}c contains data from a range of different droplet area fractions in the binding layer, so that droplet collision kinetics does not visibly influence valence.}
To understand valence selection theoretically, we therefore augment the free energy functional ${\cal F}_p$ for patches of $n$ molecules with entropic contributions describing the partition of the total number $N$ of molecules on a droplet between patches and the exterior.

Let us examine the transition of valence $Z=1$ to $Z=2$ for a droplet with $N$ total molecules exposed to two droplets with $N^\prime$ complementary molecules, as depicted in Fig.\,\ref{fig:valence}e. 
The relevant free energy difference is between a state of $n+1$ molecules in one patch (with one unbound $A^\prime$ droplet) and a state with $n$ and $1$ molecules in two binding patches. 
If the latter state has a lower free energy, a spontaneous transition $Z=1\to 2$ will happen. 
Thus, we compute
\begin{equation}
\Delta {\cal F} = \Delta {\cal F}_p + {\cal F}_C^{(1)}-{\cal F}_C^{(2)}\,,
\label{deltafvalence}
\end{equation}
where $\Delta {\cal F}_p={\cal F}_p(n+1)-{\cal F}_p(n)-{\cal F}_p(1)$ is the difference of patch free energies and the energies ${\cal F}_C^{(Z)}$ are based on configurational entropies, evaluated through counting microstate multiplicities $\Omega^{(Z)} (n,N, N^\prime$), {\em i.e.}, 
\begin{equation}
\Delta {\cal F}_C = {\cal F}_C^{(1)}-{\cal F}_C^{(2)} = 
%{\cal F}_C^{(Z)}=
-\log (\Omega^{(1)}/\Omega^{(2)})\,.
\label{omegacount}
\end{equation}
Explicit expressions for $\Omega^{(Z)}$ are given in the Methods section. 
To good approximation, the configurational entropy difference can be decomposed as
\begin{equation}
\Delta {\cal F}_C = \Delta {\cal F}_A + \Delta {\cal F}_{ext}\,,
\label{deltaFC}
\end{equation}
where $\Delta {\cal F}_A={\cal F}_A(n+1)-{\cal F}_A(n)-{\cal F}_A(1)$ is the contribution of binder molecules in the patches, {\em i.e.}, ${\cal F}_A(n)= - n \log A_p(n)$. 
The term depending on the populations of unbound molecules is
\begin{equation}
\Delta {\cal F}_{ext} (n,N, N^\prime)= \log(n+1) - N N^\prime \frac{a_{mol}A_p(1)}{A_0^2}\,.
\label{deltaFext}
\end{equation}
Here, $a_{mol}$ is the interaction area within which two DNA molecules bind and $A_0=\pi R_0^2$ is the droplet surface area. 
The last term of \eqref{deltaFext} is the leading effect of unbound molecules avoiding a binding encounter (see Methods for details). 

Comparing the free energy of a droplet having one or two droplet-droplet bonds predicts a threshold number of DNAs above which it becomes thermodynamically favorable to seed a second adhesion instead of adding binders to an existing one. 
Using as input the observed average experimental values at valence transition ($N\approx 2400$, $N^\prime\approx 27400$), the theory again has no adjustable parameters. 
We obtain a threshold value for $n$ above which $\Delta {\cal F}>0$ and valence 2 becomes favorable, namely $n_{1\to 2}\approx 430$ (Fig.~\ref{fig:valence}e). 
Both the differences in patch energies and in configurational entropies are significant factors in determining this number.  

Note that droplets with a single patch can have any number of binders in the patch below $n_{1\to 2}$. 
To compare with experiment, we assume a uniform distribution and predict a mean number of binders for $Z=1$ droplets as $\langle n\rangle_1\approx 215$. Similarly, counting microstates for higher $Z$ and averaging over all $n$ compatible with a given $Z$, we obtain the prediction $\langle n\rangle_Z \approx n_{1\to 2}(1-1/(2Z))$ (see Methods), given by the red line in Fig.~\ref{fig:valence}f. For $Z\leq 3$, these predictions are in excellent agreement with the experimental mean numbers. 

Droplets with $Z\geq 4$ are increasingly likely to be geometrically frustrated in their 2D layer (an artifact of the experiment allowing their detailed observation), i.e., the approach of additional binding partners is sterically hindered and equilibrium valence is not reached. 
In these cases, the number of binders in a patch will grow beyond $n_{Z\to Z+1}$ (open circles in Fig.~\ref{fig:valence}f), leading to supersaturated adhesion patches. 
Conservatively excluding only droplets with clearly supersaturated patches from our samples (with binder numbers exceeding the median by more than two standard deviations, see SI), the remaining droplets are characterized by unimodal $n$ distributions even for $Z=4$ and $Z=5$ (see SI), and their averages $n_4$, $n_5$ are again in very good agreement with theory (Fig.~\ref{fig:valence}f, filled symbols).
Removing the geometric constraints, e.g.\ by droplet agitation, will thus give access to higher equilibrium valence where desired.

While the average $Z$ can be adjusted by titrating the amount of DNA on the droplets, there is still a variation in $Z$ within a given bulk sample. 
Some level of variation in $Z$ is to be expected even at a thermodynamically accessible energetic minimum, as shown in Ref.~\cite{Angioletti-Uberti2014}. 
But is the variation mainly of such stochastic nature, or is it imprinted onto individual droplets because of slight differences in DNA coverage in a given population?
To answer this question, we transiently raise the temperature in experiment and track a population of individual droplets through their assembly, disassembly via melting of the bonds, and reassembly. 

\begin{figure}
  \centering
  \includegraphics{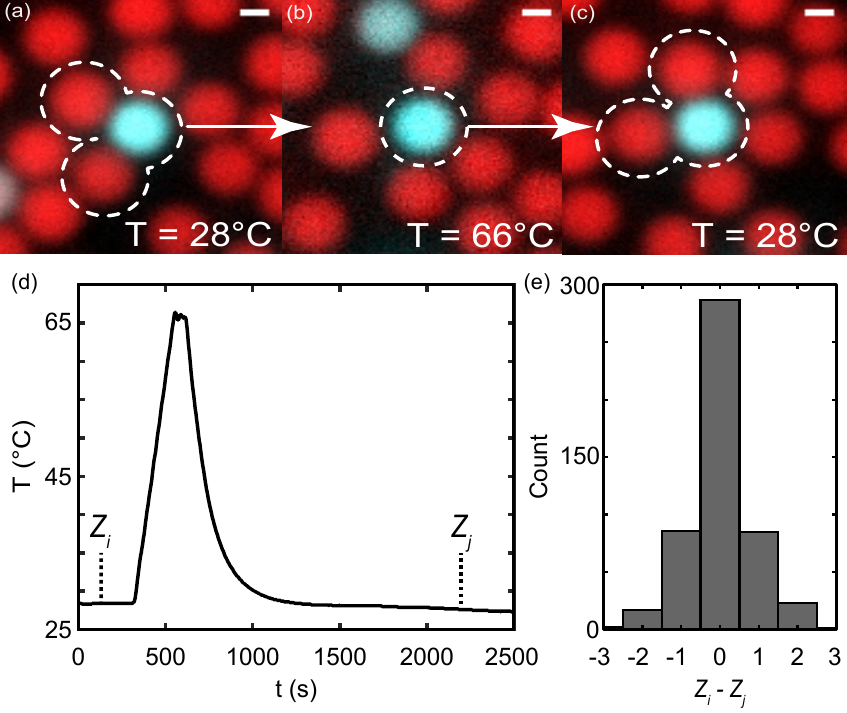} % requires the graphicx package
  \caption{Thermodynamic valence control. Fluorescent movies of a dense monolayer of complementary droplets allow us to track blue droplets before (a), during (b), and after melting of DNA bonds (c).
  %[[changed from "densely packed", because to me that implies rcp or crystal]]
  Scale bars are 2 $\mu$m. (d) After self-assembly at room temperature, a temperature pulse melts and reassembles droplet bonds on the timescale of minutes. (e) A histogram of the change in valence before and after melting shows that most droplets recover their prescribed valence despite frequent collisions with their neighbors.}
  \label{fig:valencerecover}
\end{figure}

We use a layer of droplets ($R_0 \approx 1.75 \,\mu$m) at high area fraction and a ratio of $A$ to $A^\prime$ species of 1:8 to enhance the kinetics of binding. Both flavors of droplets are internally labeled with different fluorescent dyes to distinguish them. 
The three images in Fig.\,\ref{fig:valencerecover}a-c show the same tracked $A$ droplet before melting, during melting, and upon reassembly. 
Strikingly, the droplet binds to two and only two red $A^\prime$ droplets both before and after melting, even though it has ample opportunity to bind to others. 
Figure\,\ref{fig:valencerecover}d shows a plot of the temperature protocol over time, indicating that thermodynamic valence control can be established on the order of minutes.  
Figure\,\ref{fig:valencerecover}e shows a histogram of $Z_i-Z_j$---the difference in valence before and after melting---for all the $A$ droplets observed. 
\textcolor{black}{While melting at a consistent temperature regardless of $Z_i$,} the majority of droplets do not change valence, \textcolor{black}{again independent of droplet area fraction,} allowing us to conclude that the observed spread in valence in the bulk is not caused by differences in assembly kinetics, but is due to the nonuniform coverage of droplets.
These results further support the existence of a thermodynamically stable valence, explaining the spontaneous assembly of colloidomer chains \cite{McMullen2018}. 

\section*{Discussion}

In summary, we see that it is the substantial entropic penalty of bound DNA molecules confined in a binding patch that allows for an effective binding energy on the order of $k_BT$, so that molecules inside and out of the patch can establish a meaningful thermodynamic balance. This in turn allows the system to explore the energetic effects of changing the number of patches, {\em i.e.}, valence. 
We have shown that thermodynamic valence control is a natural and expected consequence of the free energy functional, and that the values of DNA coverage where valence transitions can be expected are given by the properties of the individual binder molecule, in particular its entropic spring energy and its interaction potential. 
Thus, binder concentrations on droplets with desired valence can be tailored by the length, flexibility, and charge of the binder molecule. 
Further improvements in valence specificity could be achieved by consulting the thermodynamic model or the  implementation of simple purification procedures.
Our results also imply that valence is switchable {\it in situ} as a function of pH, salt concentration, or DNA toehold displacement reactions.
%This method of prescribing valence to isotropic droplets has the advantage that simply cycling temperature leads to bulk control, without the need to filter, segregate, or select particles that are synthesized with the correct number and geometry of patches.

The fact that droplets stay undeformed during the valence selection process plays an important role, as for a given binder, a whole spectrum of elongations and local concentrations is found in the patches and leads to unique, non-uniform optimized solutions that are not accessible in assemblies with flat, uniform patches, such as those found between liposomes. 
%These nontrivial profiles invoke a segregation of binders according to their length, sequestering DNA species in a hierarchy within the patch. [[do we need this last sentece about segregation? it seems out of place.]]

It is worthwhile stressing that valence established in this way is a function of the prepared coverage of {\em both} species of droplets. 
Just like initially isotropic atoms develop chemical valence only when encountering other atoms, and valence is specific to the intrinsic electronic properties of both binding partners, we can here program isotropic droplets for predetermined valence through the simple means of establishing their intrinsic density and type of molecular binders.

\vspace*{5mm}

{\bf Acknowledgments:} 
The authors would like to thank Frank Scheffold, Jerome Bibette, Francesco Sciortino, John Crocker, and David Pine for insightful discussions. 
This work was supported by the Materials Research Science and Engineering Center (MRSEC) program of the National Science Foundation under Grants No. NSF DMR-1420073, No. NSF PHY17-48958, and No. NSF DMR-1710163.

\vspace*{5mm}
{\bf Author Contributions:} 
A.M., S.H., and J.B. designed the research. 
A.M. performed the experiments and analyzed the data. 
S.H. formulated the theory and provided theoretical predictions.
A.M., S.H., and J.B wrote the paper. 

	\newpage

	\centerline{\bf Methods}

	\small
	\vspace*{5mm}

{\bf Droplet synthesis.} 
Monodisperse PDMS droplets were synthesized according to the protocol outlined in \cite{McMullen2018} and \cite{Zhang2017}. 
Briefly, an amount ranging from 1\% to 20\% v/v of diethoxydimethysilane (Sigma Aldrich) was dissolved in DI water and 20\% v/v ammonia. 
The droplets were then cleaned in presence of 5\,mM sodium dodecyl sulfate (SDS), either by dialysis or by centrifugation, of ammonia and reaction byproducts.
To fluorescently label droplets, we first incubated the cleaned droplets in a small amount of Azido-PEG5-triethoxysilane (BroadPharm) in 1\% v/v ammonia and 5 mM SDS. 
The droplets were then washed by centrifugation and incubated with a Cyanine3 DBCO or Cyanine5 DBCO compound (Lumiprobe). Unreacted dye was then washed out via centrifugation. 
Droplets used for droplet-droplet measurements were synthesized to be denser than water so that they would sink. 
This was done by substituting half the diethoxydimethysilane for (3,3,3-trifluoropropyl)methyldimethoxysilane (Gelest). 

{\bf DNA-labeling of emulsion droplets.}
DNA strands labeled with a reactive azide compound were purchased from Integrated DNA Technologies. 
DNA was diluted to 100 $\mu$M in 50 mM NaCl TE buffer, and then mixed in equal volume with 200 $\mu$M 1,2-distearoyl-sn-glycero-3-phosphoethanolamine-N-[dibenzocyclooctyl(polyethylene glycol)-2000] (ammonium salt) (DSPE-PEG-DBCO, Avanti Polar Lipids). 
Both complementary strands (see supplemental for all DNA strand sequences) of a single DNA complex are each attached to a hydrophobic lipid through a copper-free ring-strain promoted alkyne-azide cycloaddition reaction \cite{Wang2015,agard2004strain}.  
The resulting DNA-lipid complexes were mixed together with the complementary spacer DNA-lipid complex and annealed together, then diluted to a known concentration and incubated with a dilute droplet sample in 50 mM NaCl TE. 
Droplets were then washed with 50 mM NaCl 0.1\% Triton x-100 solution once. 
Triton x-100 was empirically found to wash off any loosely attached or single stranded DNA compounds. 
The labeled droplets were then washed several times in 50 mM NaCl to remove the Triton x-100 and stored in 50 mM NaCl. 
DNA with a biotin-streptavidin chemistry was labeled according to a different procedure outlined in ref \cite{Zhang2017}.
Only the longest DNA strands in Fig.\,\ref{fig:AvsI}(d) were attached using this chemistry. 

{\bf DNA-labeling of glass surfaces}
First, glass channels are created using coverslips and UV glue.
Briefly, two glass coverslips are glued to a microscope slide using UV glue to create an open channel.
A third coverslip is then glued on top to create an open channel.
These are then cleaned with KOH saturated IPA (base bath), O$_2$ plasma cleaned for 30 minutes, and then hydrophobized with hexamethyldisilazane (\textit{Sigma Aldrich}). 
A 1 $\mu$M solution of the DNA complex complementary to that on the droplets (the A DNA) is added to the glass channel and incubated for thirty minutes before flushing with buffer to 
remove free DNA.
% We observed the binding of DNA-labeled emulsion droplets to adhesive surfaces using brightfield and fluorescence imaging.
% Brightfield was used to locate droplets adhered to the surface. 
The DNA coating the glass surface is tagged with a cy5 cyanine dye, revealing the DNA-labeled glass surface and the surface portion of the adhesion patch. 
% The DNA coating the droplet is tagged with a cy3 cyanine dye. 
% The signal from the cy3 dye is entirely located in the droplet portion of the adhesion patch in most experimentally relevant conditions.

{\bf Measurement of DNA coating density on droplets}
The number of DNA on the droplets was measured following the method used in Ref.~\cite{Zhang2017}.
Briefly, for each condition, a known quantity of DNA was added to a known number of droplets. 
The fluorescence intensity of the buffer containing the DNA was measured with a Horiba PTI QuantaMaster fluorimeter prior to the addition of droplets. 
After labeling the droplets with DNA, the droplets were washed and the wash buffer was retained. 
The intensity of the wash buffer was then measured, and the number of DNA's that went onto the droplets was estimated by the difference between these intensity measurements. 
We found that under almost all conditions $\geq 90\%$ of DNA went onto the droplets.

{\bf Droplet-surface adhesion measurements and analysis}
Droplets were added to a custom-built flow cell containing a DNA-labeled surface in a buffer containing 20 mM MgCl$_2$, 0.1 \% w/v Brij-35 surfactant, and 5 mM Tris buffer at pH 8.
Images are taken with a 100x oil immersion lens. 
First a brightfield image is taken, then fluorescence images are taken of the cy3 (A') and cy5 (A) dyes, respectively. 
Using custom MATLAB software, the droplets are first located using the brightfield image. 
Then, the adhesion patches on the cy3 and cy5 channel are located within each droplet using a Hough transform. 
The Hough transform identifies the centroid and radius of the adhesion patch. 
The intensity per pixel of the adhesion patch is calculated by summing the intensity of all pixels inside the adhesion patch and dividing by the number of pixels inside the patch.
To calculate the radial profile of the adhesion patch, rays are drawn every 15 degrees from the center of the patch to a distance three times the patch radius. 
The intensity along this arc is interpolated from the values of the neighboring pixels. 
The resulting profile for each ray is then averaged, resulting in a measurement of the radial profile of the adhesion patch.

{\bf Droplet-droplet adhesion measurements and analysis}
Droplets were prepared according to the above methods and then added together in a custom built flow cell.
The $A^\prime$ species was introduced at an approximately 8 to 1 number ratio with the $A$ species. 
The amount of DNA on the $A$ species was varied among different conditions.
The buffer contained 20 mM MgCl$_2$, 0.5 w/v \% F38 pluronic surfactant, and 5 mM Tris buffer at pH 8. 
Images are taken with a 100x oil immersion lens. 
A brightfield image is taken, and then fluorescence images are taken of the cy3 and cy5 dyes. 
Using custom MATLAB software, the droplets are first located using the brightfield image. 
The fluorescence images are used to identify which droplets are $A$ and which are $A^\prime$.
For each $A$ droplet, the perimeter intensity is measured. 
Patches are located by detecting sharp changes in fluorescence intensity.
Their size is measured by extrapolation from their arc length, with the implicit assumption that they are symmetrical. 
The valence is measured by counting the number of detectable patches. 
The total DNA $N$ per droplet is estimated by integrating the total intensity on the droplet,  both inside and outside the patches, and setting the average total intensity of each batch to the value obtained by fluorimetry measurements.
Alternatively, valence measurements were also obtained from tracking particles 
and assigning valence based on the persistence of connection, resulting in valences consistent with the fluorescence measurements. 

{\bf Valence recovery measurements}
Droplets were prepared according to the above methods and then added together in a custom built flow cell attached to a temperature controller.
The $A^\prime$ species, labeled internally with cy5, was introduced at an approximately 8 to 1 number ratio with the $A$ species, labeled internally with cy3. 
The amount of DNA on the $A$ and $A^\prime$ species was sufficient for an average valence of about 2.
Note that we observed a slightly larger variation in measured valence in internally dyed particles, which could be attributed to a change in surface chemistry from the presence of the dye.
The buffer contained 20 mM MgCl$_2$, 0.05 w/v \% F68 pluronic surfactant, and 5 mM Tris buffer at pH 8. 
% Images are taken with a 60x oil extra long working distance air objective. 
The $A$ droplets were tracked through the course of several melting and reassembly cycles. 
Droplets were given approximately 30 minutes to reform connections before the next cycle began. 
The valence was measured by tracking the number of connected $A^\prime$ particles both before and after a temperature pulse to 65$^\circ$C, which melted all connected bonds.
Results were checked by hand to ensure accuracy. 
We confirmed that enough time elapsed between the melting and reforming of bonds that the particles were not just rebinding to their previous partners.

\vspace*{5mm}

{\bf Molecular interaction energy.} 
A link between two surfaces consists of the semiflexible dsDNA binder (unless stated otherwise, we use a 50 bp complex with contour length $L_D\approx 40$\,nm and persistence length $\xi_D\approx 50$\,nm) with elastic coils of PEG molecules attached at both ends (at molecular weight 2000, 
unstretched coils have radius $R_P\approx 2.4$\,nm and contour length $L_P\approx 15$\,nm). 
Interaction between two such binders spanning surfaces of distance $h$ is 
modeled as Onsager repulsion of rods, taking into account that for $h>L=(L_D+2R_P)$ there is still excluded volume between the parallel rods governed by the width length scale $d_D$, where $d_D\approx 3$\,nm is the effective width of the DNA molecule at the experimental salt concentration \cite{hsieh2008ionic}. We thus arrive at an interaction energy (in units of $k_BT$) of \cite{onsager1949effects}
\begin{equation}
g(z)=2(L+d_D -h)/L\,.  
\label{gofz}
\end{equation}
While the PEG coils could in principle stretch to their contour length resulting in a total binder length of $L_D+2L_P$, this would require much more energy than is accessible throughout the experimental range of parameters. Thus, the length of PEG molecules remains close to $R_P$ and rod-like interaction is governed by the length scale $L$. 
%In practice, the PEG coils are never significantly stretched for the experimentally accessible ranges of $n$
%where the total contour length is $L=L_D+2 L_P$, as given in the main text. 
%acknowledging that the effective rod length of the binder is the DNA length plus twice the size $r_P=R_P/L$ of the (unextended) PEG coils, where $R_P\approx 2.4$\,nm. 

\vspace*{5mm}

{\bf Molecular spring energy.} 
Each composite binder is a series of three molecular springs, two PEG coils and the DNA complex. Thus, the molecular spring energy is
\begin{equation}
s(h)= s_D(h_D) + 2s_P(h_P) \,,
\label{stot}
\end{equation}
where $s_D$ is the DNA spring energy, modeled as a semi-flexible nonlinear spring with finite extensibility, and $s_P$ is the spring energy of a PEG coil well approximated by an ideal chain for the range of small extensions considered in this work. The total surface-to-surface distance is covered by DNA and PEG, i.e., $h=h_D+2h_P$.

A PEG spring has a Gaussian coil energy of
 \begin{equation}
 s_P(h)= \sigma_P (h_P(h)-R_P)^2\,
 \label{sp}
 \end{equation}
 when taking up a distance $h_P$ between surfaces. Here, $\sigma_P=3/(4\xi_P L_P)$, with the persistence length $\xi_P\approx 0.38$\,nm.
 
The dsDNA is a semiflexible nonlinear spring with finite extensibility ($L_D\sim \xi_D$). We use the approximation by Winkler \cite{winkler2003deformation},
 \begin{equation}
 s_D(h)= \sigma_D  \left(1-\frac{(h-2h_P(h))^2}{L_D^2}\right)^{-1}\,,
\label{sd}
 \end{equation}
 with the prefactor $\sigma_D=3L_D/(4\xi_D)$. 
To specify the relative extent of DNA and PEG (i.e., $h_P(h)$), we use the condition of equal forces in the PEG-DNA-PEG series of springs, namely,
\begin{equation}
\frac{\partial s_P}{\partial h_P}= \frac{\partial s_D}{\partial h_P}\,.
\label{fs}
\end{equation}
The resulting $h_P(h)$ is a lengthy but explicit expression and allows evaluation of $s(h)$ (see Supplementary Information).

\vspace*{5mm}

{\bf Determining concentration profiles in a patch.}
The energy functional of a binding patch containing a given number $n$ of 
linkers is described by ${\cal F}_p={\cal F}_S+{\cal F}_{int}$ -- we omit the constant binding energy. Written explicitly for a circular patch, we have
\begin{equation}
{\cal F}_p =  \int_0^R \left[  c(r) s(r) + (c^2(r)/c_m) g(r) + \Lambda c(r) \right] (2\pi r)dr  - n\Lambda\,,
\label{fpexpl}
\end{equation}
where $c(r)$ is the area density of linkers and we have introduced a Lagrange multiplier $\Lambda$ to enforce the total number $n$.

Minimizing (\ref{fpexpl}) with respect to $c(r)$ and enforcing $n=\int c(r) 
(2\pi r)dr$, we arrive at (\ref{cexplicit}) with (\ref{gammadef}). Setting $c(r_p)=0$ readily yields (\ref{rpeq}) as an implicit equation for $r_p(n)$. An explicit expression for ${\cal F}_p$ as a function of $n$ is obtained by plugging these results back into \eqref{fpexpl} (see SI).
%Consistent with the expectation from the geometry of an undeformed spherical droplet, the limit of $n\to\infty$ results in a patch packed with DNA binders up to the maximum radius $r_{p,max}=\sqrt{2R_0 L}$. 
%In practice, the PEG coils are never significantly stretched for the experimentally accessible ranges of $n$, so that the plateau value of $r_p$ is close to $\sqrt{2R_0 L}$ instead, cf.\ Fig.~\ref{fig:profiles}e.

For direct comparison with experiment, the measured fluorescence intensities are first calibrated as follows: For a batch of droplets incubated at 
the highest amount of DNA complex, we measure the difference of DNA in the fluid bulk to determine the average amount of DNA attached to each droplet, i.e., the $\langle N^\prime\rangle$ value. When these droplets are bound to 
a substrate, the overwhelming majority of binders crowd into the patch ($n\approx N^\prime$), so that the measured patch fluorescence intensity corresponds to a mean concentration $\bar{c}_p=n/(\pi r_p^2)$, where $r_p$ is measured. All concentration values in the present work use this calibration. Choosing experimental profiles with closely matching $\bar{c}_p$, a theoretical patch radius is determined from the measured $r_p$ taking into account
the convolution with the fluorescence microscope's PSF (width 300\,nm). Via (\ref{rpeq}) this determines a theoretical value for 
$n$, and the theory profile can be plotted through (\ref{cexplicit}) without 
any free parameters. This procedure yields excellent results both for monotonic profiles at low concentrations (Fig.~\ref{fig:AvsI}f) and for the ring-shaped intensity distributions at higher concentrations (Fig.~\ref{fig:AvsI}g).

\vspace*{5mm}

{\bf Droplet-droplet binding.} 
As we assume the same undeformed, spherical droplet geometry for droplet-droplet binding as for the binding of a droplet to a flat substrate, the theory describing the latter is easily modified to describe the former by 
halving $L$. The configurations of binders to both sides of the symmetry plane between droplets are then equivalent to those of the droplet-substrate binding case, and thus the formulas derived for that case can be used without further modification.

\vspace*{5mm}

{\bf Entropic energy contributions in bound droplets.} 
To count microstates $\Omega^{(1)}$ and $\Omega^{(2)}$ for the two configurations of Fig.~\ref{fig:valence}d, we follow \cite{Parolini2015} with the necessary modifications (e.g., our droplets carry only one DNA species 
each). For $Z=1$, we have $n+1$ molecules from an $A$ and an $A^\prime$ 
droplet approaching to within a molecular scale $a_{mol}$ (see below) in order to bind in a patch. The remaining $N-n-1$ unbound molecules on the $A$ droplet are free to diffuse on its surface, but cannot approach $A^\prime$ molecules in the patch to within $a_{mol}$ (otherwise they would be bound). Taking this restriction of available states into account, we obtain the total number of microstates as
\begin{align}
\Omega^{(1)} = \binom{N}{n+1}& \binom{N^\prime}{n+1} (n+1)!  A_p^{n+1}(n+1)a_{mol}^{n+1}\times \nonumber\\ \times A_0^{N^\prime} A_0^{N^\prime-n-1} &\left(A_0-a_{mol} N^\prime \frac{A_p(n+1)}{A_0}\right)^{N-n-1}\,,
\label{om1}    
\end{align}
where $A_0=4\pi R_0^2$ is the droplet surface area. Likewise, the $Z=2$ configuration yields
\begin{align}
\Omega^{(2)} =  \binom{N}{n}&\binom{N^\prime}{n}\binom{N-n}{1}\binom{N^\prime}{1}
 n! A_p^{n}(n) A_p(1) a_{mol}^{n+1}\times \nonumber\\ \times A_0^{N^\prime-n} A_0^{N^\prime-1} &\left(A_0-a_{mol} N^\prime \frac{A_p(n)+A_p(1)}{A_0}\right)^{N-n-1}\,.
\label{om2}    
\end{align}
We use these expressions in \eqref{omegacount} and observe that the approximations 
$n\ll N,N^\prime$, $a_{mol}(N,N^\prime)A_p(n)/A_0\ll 1$, and $A_p(n+1)-A_p(n)\ll A_p(1)$ hold to great accuracy for the range of parameters considered here. Taylor expansions in small arguments then result in \eqref{deltaFext} to leading order; more explicit details are given in the SI.

To obtain an approximation for $a_{mol}$, we follow the theory for binders of bound length $L=L_D+2R_P$ with sticky ends \cite{mognetti2012controlling,Parolini2015} between surfaces a distance $h$ apart. Binders can stick to complementary partners for $h\leq \tilde{L}$, where $\tilde{L}=L+\ell_s$ using the sticky-end length $\ell_s$ Averaging over all available $h\in[0,\tilde{L}]$ obtains $a_{mol}=\frac{2}{3}\pi \tilde{L}^2$, closing the formalism.

Performing microstate counts for higher $Z$ results, to leading order in these approximations, in the same ratio of $\Omega^{(Z)}/\Omega^{(Z+1)}$, and thus the same nominal threshold coverage per patch, $n_{1\to 2}$. However, a valence $Z$ droplet compatible with these theoretical thresholds can have anywhere between $(Z-1)n_{1\to 2}$ and $Z n_{1\to 2}$ binders in all $Z$ patches. Assuming a uniform distribution of $n$ between these bounds gives the prediction of average coverage per patch,
\begin{equation}
\langle n\rangle_Z = n_{1\to 2} \left(1-\frac{1}{2Z}\right)\,,
    \label{nofztheo}
\end{equation}
displayed as a red line in Fig.~\ref{fig:valence}f.

% \bibliographystyle{apsrev4-1}
% \bibliography{bibliography_v1.bib}

%

\end{document}

\maketitle

\section{DNA Sequences}
The following is a complete list of DNA sequences used in this work, listed with their modifications from 5$^\prime$ to 3$^\prime$.
\begin{itemize}
    \item Azide Cy3A GCA TTA CTT TCC GTC CCG AGA GAC CTA ACT GAC ACG CTT CCC ATC GCT A  GA GTT CAC AAG AGT TCA CAA 
    \item Azide Cy5 A GCA TTA CTT TCC GTC CCG AGA GAC CTA ACT GAC ACG CTT CCC ATC GCT A  TT GTG AAC TCT TGT GAA CTC
    \item TAG CGA TGG GAA GCG TGT CAG TTA GGT CTC TCG GGA CGG AAA GTA ATG CT Azide
    \item Azide A GCA TTA CTT TCC GTC CCG A GA GTT CAC AAG AGT TCA CAA
    \item AzideN iCy3 A GCA TTA CTT TCC GTC CCG A TT GTG AAC TCT TGT GAA CTC
    \item  TCG GGA CGG AAA GTA ATG CT Azide
    \item Biotin TC TTG CTG GAA TCC TAA GTG GAG CTC ACG TAT CAT CGA ACT CGT CAA CAC TAT ATA AGA TTA AGT TAG TTG AAA GAA CTA CAC TAG GAA CAG ACC GTC CTC TTG TCT AGT TAT GGA ATT TAT ATT GAG AAG GTC TAG TAA ATT GTA GTC TAC GAG TCA ATT AAT GAC GCA ACC CAA TCC AGA GTT CAC AAG AGT TCA CAA
    \item Biotin TC TTG CTG GAA TCC TAA GTG GAG CTC ACG TAT CAT CGA ACT CGT CAA CAC TAT ATA AGA TTA AGT TAG TTG AAA GAA CTA CAC TAG GAA CAG ACC GTC CTC TTG TCT AGT TAT GGA ATT TAT ATT GAG AAG GTC TAG TAA ATT GTA GTC TAC GAG TCA ATT AAT GAC GCA ACC CAA TCC ATT GTG AAC TCT TGT GAA CTC
    \item TGG ATT GGG TTG CGT CAT TAA TTG ACT CGT AGA CTA CAA TTT ACT AGA CCT TCT CAA TAT AAA TTC CAT AAC TAG ACA AGA GGA CGG TCT GTT CCT AGT GTA GTT CTT TCA ACT AAC TTA ATC TTA TAT AGT GTT GAC GAG TTC GAT GAT ACG TGA GCT CCA CTT AGG ATT CCA GCA AGA
\end{itemize}

\section{Supplemental Movie 1}
This movie shows an A droplet (cyan) surrounded by A$^\prime$ droplets (red).
The A droplet is tracked through the entire movie. 
The movie is shown in the A droplets frame of reference, so that it does not appear to diffuse. 
The A droplet starts bound to two A$^\prime$ droplets. 
The temperature, shown in degrees C in the upper right corner, is increased and the bound droplets come off and diffuse away. 
The temperature is turned back down to room temperature, and the A droplet finds two new A$^\prime$ droplets to bind to. 
 
\section{Patch free energy functional}
We here detail the systematic derivation of the free energy functional for a circular binding patch with $n$ binders, with binder concentration $c(r)$ and dimensionless spring and interaction energies $s(r)$ and $g(r)$, respectively. If every binder has effective binding energy $\epsilon$, this makes a constant contribution $-n\epsilon$ to the functional, which can be disregarded for optimization purposes. We enforce the condition of $n$ binders via an integral constraint with Lagrange multiplier $\Lambda$, writing 
\begin{equation}
    {\cal F}_p = {\cal F}_{s} + {\cal F}_{int} + {\cal F}_\Lambda\,,
\end{equation}
where
\begin{equation}
{\cal F}_{\Lambda} = \Lambda\left( \int_0^{r_p} c(r) (2\pi r)dr - n \right)\,.
\end{equation}
Thus,
\begin{equation}
{\cal F}_p =  \int_0^{r_p} \left[c(r) s(r) + (c^2(r)/c_m) g(r) + \Lambda c(r) \right] (2\pi r)dr  - n\Lambda\,.
\label{fptot}
\end{equation}
Minimization with respect to $c$ is straightforward and yields 
\begin{equation}
c(r) = \frac{c_m}{2}\frac{-\Lambda-s(r)}{g(r)} \,.
\label{c1}
\end{equation}
Note that the molecular mechanics functions $s$ and $g$ are really functions of surface-to-surface distance, which is translated to radial dependence via the shape of the interfaces.
Next we obtain $\Lambda$ from the number constraint,
\begin{equation}
n = \int_0^{r_p} \left[ \frac{c_m}{2}\frac{\epsilon-\Lambda-s(r)}{g(r)} \right](2\pi r)dr \,,
\end{equation}
giving
\begin{equation}
\Lambda=-\frac{2N}{c_m\pi R^2 \Gamma_0} - \frac{\Gamma_1}{\Gamma_0}\,,
\label{Lamb}
\end{equation}
where we have defined the dimensionless functions
\begin{equation}
\Gamma_k \equiv \Gamma_k(r_p)= \frac{1}{\pi r_p^2} \int_0^{r_p} \frac{s^k(r)}{g(r)} (2\pi r)dr\,.
\end{equation}
Plugging $\Lambda$ back into $c(r)$ gives the explicit concentration profile
\begin{equation}
c(r) = \frac{c_m}{2g(r)}\left(\frac{2n}{c_m \pi r_p^2 \Gamma_0} + \frac{\Gamma_1}{\Gamma_0} -s(r)\right) \,.
\label{cexplicit}
\end{equation}
The radius $r_p(n)$ of the edge of the patch is then given by $c(r_p)=0$, i.e., 
\begin{equation}
\frac{2n}{c_m\pi r_p^2} = s(r_p) \Gamma_0 - \Gamma_1\,.
\label{patchsize}
\end{equation}
Now using (\ref{Lamb}), (\ref{cexplicit}) in (\ref{fptot}) yields the patch free energy as a function of $n$,
\begin{equation}
{\cal F}_p(n) = \frac{n}{\Gamma_0}\left(\frac{n}{c_m\pi r_p(n)^2}+\Gamma_1\right) + \frac{c_m\pi r_p(n)^2}{4}\left(\frac{\Gamma_1^2}{\Gamma_0}-\Gamma_2\right) \,.
\label{fpofn}
\end{equation}

\section{Spring energy of the binder composite}
\begin{figure}[h]
  \centering
  \includegraphics[width=12cm]{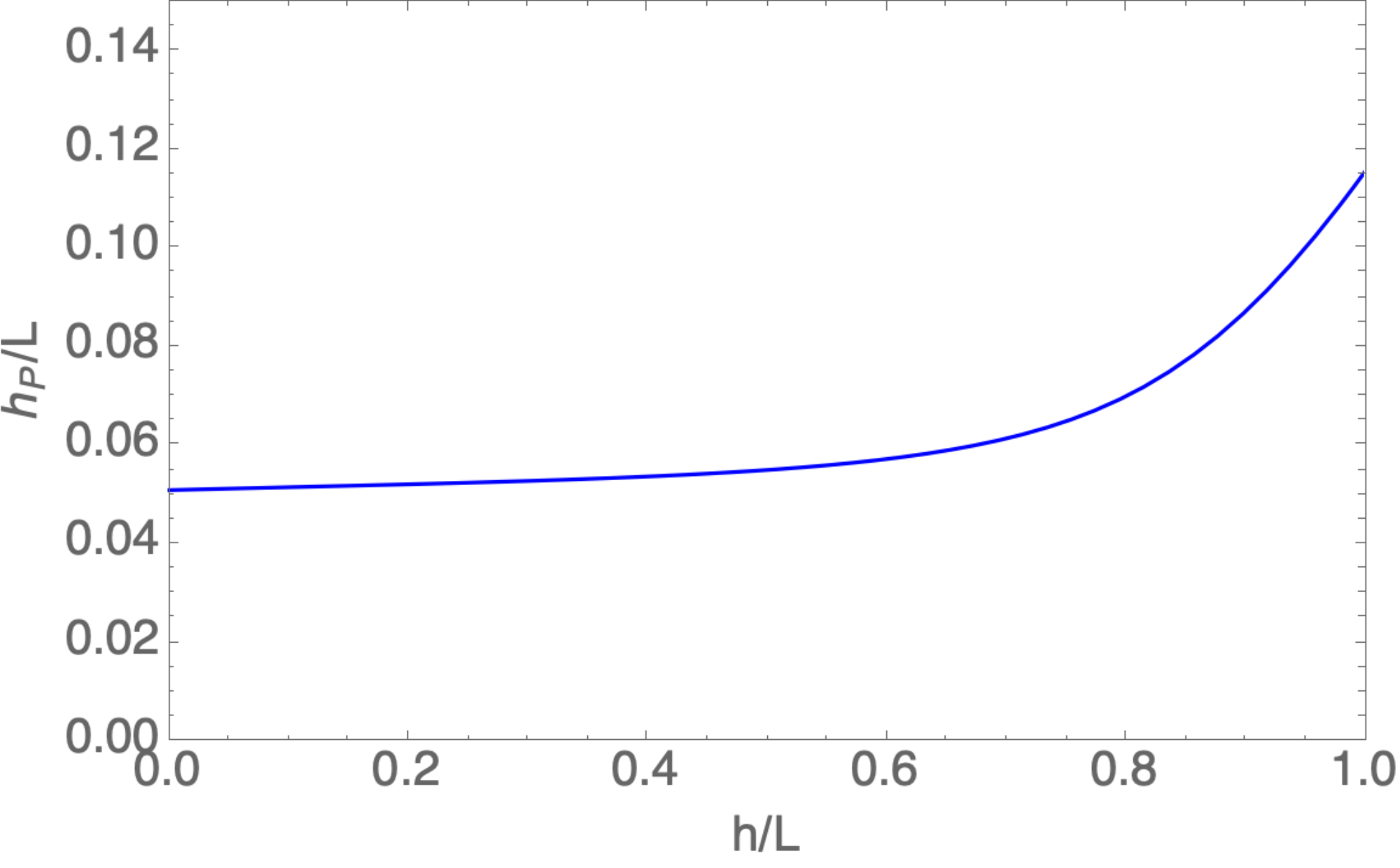} % requires the graphicx package
  \caption{Stretch length $h_P$ of one PEG coil as a function of the total distance $h$ between interfaces bridged by the dsDNA and the two PEG molecules bracketing it. Both $h$ and $h_P$ are normalized by the approximate maximum length $L=L_D+2R_P$.}
  \label{fig:sihp}
\end{figure}

After bonding the sticky ends of an $A$ and an $A^\prime$ molecule, the resulting binder is a chain of three springs, a dsDNA semiflexible polymer framed by two PEG coils. As given in the methods section, the total energy is then 
\begin{equation}
s(h)= s_D(h_D) + 2s_P(h_P) \,,
\label{stot}
\end{equation}
with
 \begin{equation}
 s_P(h)= \sigma_P (h_P-R_P)^2\,,
 \label{sp}
 \end{equation}
 a Gaussian spring with $\sigma_P=3/(4\xi_P L_P)$ taking up a distance $h_P$ between surfaces and being in equilibrium at a Gaussian coil extent $R_p$. The dsDNA is taking up the remaining $h-2h_P$ distance (the total distance between surfaces being $h$, and assuming symmetric PEG coils). Thus, \begin{equation}
 s_D(h)= \sigma_D  \left(1-\frac{(h-2h_P)^2}{L_D^2}\right)^{-1}\,,
\label{sd}
 \end{equation}
 with the prefactor $\sigma_D=3L_D/(4\xi_D)$. 
 Springs in series must have equal load, so we enforce (with $h_P$ as the variable)
 \begin{equation}
\frac{\partial s_P}{\partial h_P}= \frac{\partial s_D}{\partial h_P}\,.
\label{fs}
\end{equation}
which represents a condition for $h_P(h)$. The physical solution of (\ref{fs}) with (\ref{sp}), (\ref{sd}) is the largest positive root of a fifth-order polynomial,
\begin{equation}
\sum_{m=0}^{5} \alpha_m h_p^m = 0\,.    
\end{equation}
Abbreviating $L_p-R_p\equiv \ell_p$, we have
\begin{align}
\alpha_0 &= \frac{2\sigma_D h}{L_D^2} + \frac{\sigma_P R_P}{\ell_P^2} + \frac{\sigma_P R_P h^4}{\ell_P^2L_D^4} - \frac{2\sigma_P R_P h^2}{\ell_P^2L_D^2}\,,\\
\alpha_1 &= -\frac{4\sigma_D}{L_D^2}-\frac{\sigma_P}{\ell_P^2}-\frac{8\sigma_P R_P h^3}{\ell_P^2L_D^4}-\frac{\sigma_Ph^4}{\ell_P^2 L_D^4} + \frac{2\sigma_P h^2}{\ell_P^2L_D^2} + \frac{8\sigma_P R_P h}{\ell_P^2L_D^2}\,,\\
\alpha_2 &= \frac{24\sigma_P R_P h^2}{\ell_P^2 L_D^4}+\frac{8\sigma_P h^3}{\ell_P^2 L_D^4} - \frac{8\sigma_P R_P}{\ell_P^2 L_D^2} - \frac{8\sigma_P h}{\ell_P^2 L_D^2}\,,\\
\alpha_3 &= -\frac{32\sigma_P R_P h}{\ell_P^2 L_D^4} - \frac{24\sigma_P h^2}{\ell_P^2 L_D^4} + \frac{8\sigma_P}{\ell_P^2 L_D^2}\,,\\
\alpha_4 &= \frac{16\sigma_P R_P}{\ell_P^2 L_D^4} + \frac{32\sigma_P h}{\ell_P^2 L_D^4}\,,\\
\alpha_5 &= -\frac{16\sigma_P}{\ell_P^2 L_D^4}\,.
\end{align}
The resulting $h_P(h)$ remains very close to $R_P$ for most distances $h$; only when the dsDNA is nearly entirely stretched to $L_D$ do the PEG springs engage and start stretching, as demonstrated in Figure~\ref{fig:sihp}. For the valence control regime described in the main paper, the PEG coils can be treated as unextended beyond $R_P$.

\section{Entropic free energy}
We shall count microstates for the $Z=1$ configuration illustrated in Fig.~3(d) of the main text, where an $A$ droplet with $N$ molecules is bound to an $A^\prime$ droplet with $N^\prime$ molecules, while a second identical droplet is unbound. We have to choose $n+1$ molecules out of $N$ and $N^\prime$ for binding, which can be permuted in $(n+1)!$ ways, locate these molecules in a patch of size $A_p(n+1)=\pi r_p^2(n+1)$ (provided by theory through \eqref{patchsize}) and encounter partners within the characteristic molecular area $a_{mol}$. Furthermore, there are $N^\prime$ molecules on the $A_0$ area of the unbound $A^\prime$ droplet and $N^\prime-n-1$ molecules on the binding partner $A^\prime$ droplet; these latter can roam the area $A_0$, but must not bind to any of the other $N-n-1$ molecules on the $A$ droplet, i.e., these  $N-n-1$ molecules are restricted to $A_0- a_{mol} N^\prime \frac{A_{n+1}}{A_0}$, taking the probability of the unbound $A^prime$ molecules to be in the binding patch as $\frac{A_{n+1}}{A_0}$. All together, we obtain
\begin{equation}
\Omega^{(1)} = \binom{N}{n+1} \binom{N^\prime}{n+1} (n+1)!  A_p^{n+1}(n+1)a_{mol}^{n+1} A_0^{N^\prime} A_0^{N^\prime-n-1} \left(A_0-a_{mol} N^\prime \frac{A_p(n+1)}{A_0}\right)^{N-n-1}\,,
\label{om1full}        
\end{equation}
and analogous considerations for the $Z=2$ scenario give
\begin{equation}
\Omega^{(2)} =  \binom{N}{n} \binom{N^\prime}{n}\binom{N-n}{1}\binom{N^\prime}{1}
 n! A_p^{n}(n) A_p(1) a_{mol}^{n+1} A_0^{N^\prime-n} A_0^{N^\prime-1} \left(A_0-a_{mol} N^\prime \frac{A_p(n)+A_p(1)}{A_0}\right)^{N-n-1}\,.
\label{om2full}    
\end{equation}
Thus, we obtain the ratio
\begin{equation}
\frac{\Omega^{(1)} }{\Omega^{(2)} }    = \frac{N^\prime - n}{(n+1) N^\prime}\frac{(A_p(n+1))^{n+1}}{(A_p(n))^n A_p(1)} \left[1-a_{mol}N^\prime \left(\frac{A_p(n+1)-A_p(n)-A_p(1)}{A_0^2}\right)\right]^{N-n-1}\,.
\label{ratio}
\end{equation}
Taking the negative logarithm of \eqref{ratio} yields several additive contributions to $\Delta {\cal F}_C$. The contribution $-\log(1-n/N^\prime)$ is very small for our situation of $n\ll N^\prime$ and can be neglected. The explicit combinatorial term $\log(n+1)$ is clearly visible (favoring two patches because any molecule can start a new patch). The area ratio term straightforwardly yields $\Delta {\cal F}_A$. In taking the logarithm of the last term in \eqref{ratio}, we again approximate $N-n-1$ by $N$ because $n\ll N$ and recognize that the magnitude of the factor subtracted from 1 is much smaller than 1 (so we can replace the logarithm by this small argument). Finally, we observe that for $n$ of relevant size (at least several 100 for valence control in these experiments), the patch area difference $A_p(n+1)-A_p(n)$ is very small compared to $A_p(1)$, so we neglect the former. Together, these approximations for the last term (which introduce only very small relative errors compared to the full formula) lead to the expression given in the main text,
\begin{equation}
\Delta {\cal F}_{ext} (n,N, N^\prime)= \log(n+1) - N N^\prime \frac{a_{mol}A_p(1)}{A_0^2}\,.
\label{deltaFextSI}
\end{equation}
A combinatorial factor of 2 could have been introduced choosing one of the droplets to be the binding partner for $Z=1$, but it has been omitted because the same factor chooses the patch with $n$ molecules in the $Z=2$ case.

For the valence transition $Z=2$ to $Z=3$, the same combinatorics applies, comparing a situation with two patches of $n$ and $n+1$ binders to one with two $n$ patches and one single-molecule patch. Employing all the same approximations, this computation yields the identical result (\ref{deltaFextSI}). Quantitative changes will be expected when the sum of all bound molecule numbers does not remain small compared with $N$ or $N^\prime$, \textcolor{black}{or more generally if the number of bound and unbound molecules on the droplet surface becomes very large. Then, the molecular area $a_{mol}$ should become itself $N$-dependent, and the second term of (\ref{deltaFextSI}) will show non-linear behavior and saturate.} However, for the first three transitions (modeled in the main text) our approximations are still fulfilled to very good accuracy.

Setting $\Delta {\cal F} =0$ yields the transition patch coverage $n_{Z\to Z+1}\equiv n_c$, above which a spontaneous increase of valence is expected. Bound droplets with fewer bound molecules will also have valence $Z$, so that $n_{Z\to Z+1}$ is an upper bound. The predictions of mean number of molecules per patch$\langle n\rangle$ at valence $Z$ assumes that all possible molecule numbers consistent with $Z$ are equally likely, i.e., all total binder numbers between $Z n_c$ and $(Z+1)n_c$. Thus, we predict $\langle n\rangle(Z) = n_c (1-1/(2Z))$, seen as the red line in Fig.~3f of the main text.

\section{Supersaturated patches}

\begin{figure}[h]
  \centering
  \includegraphics[width=12cm]{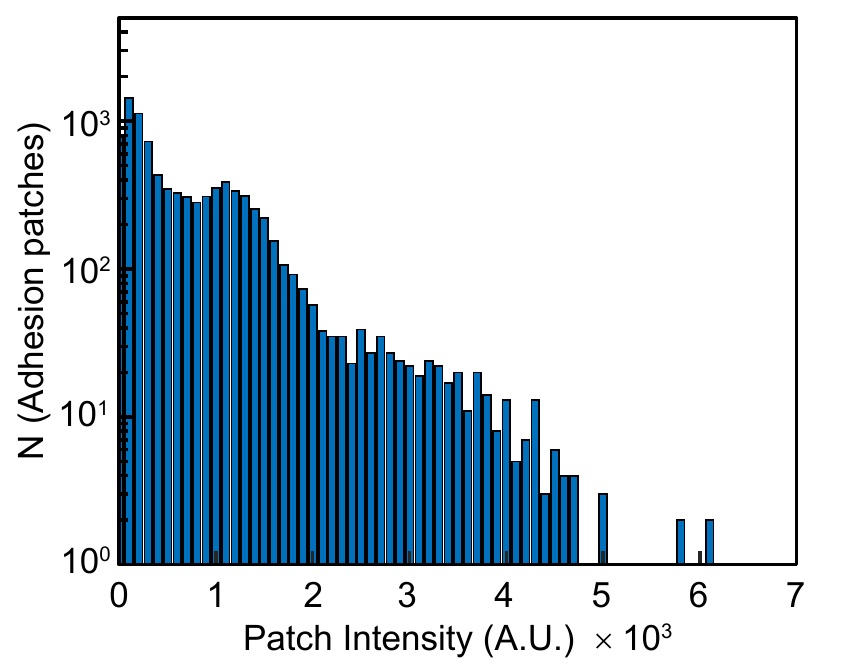} % requires the graphicx package
  \caption{Semi-log distribution of the intensity for all patches before droplets with supersaturated patches are removed. The distribution has a long tail of very bright patches which have out sized influence over calculated values.}
  \label{fig:histograms_I}
\end{figure}

\begin{figure}[h]
  \centering
  \includegraphics[width=12cm]{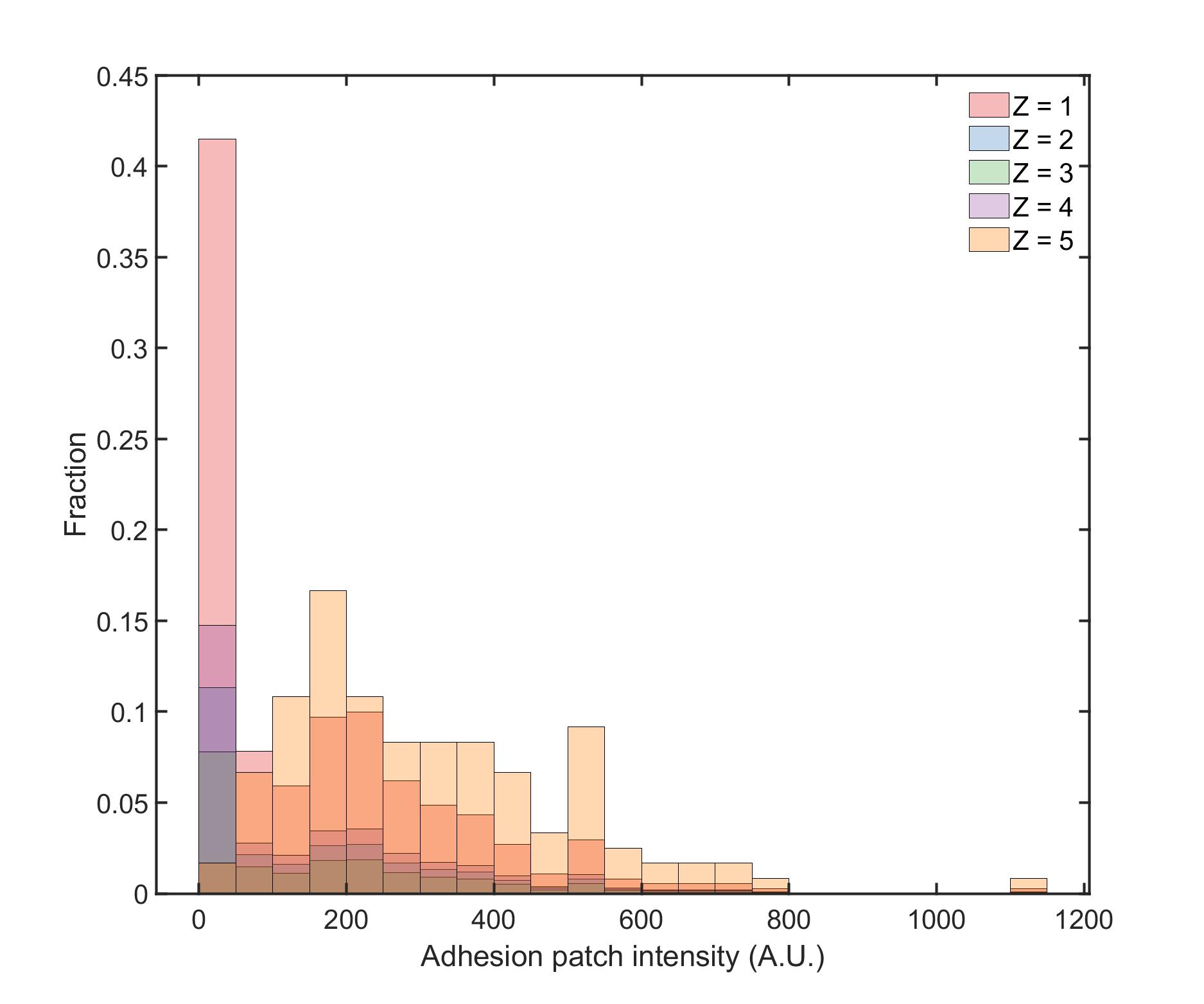} % requires the graphicx package
  \caption{Distributions of intensity for all patches for droplets with valence $Z = 1, 2, 3, 4, 5$ after droplets with supersaturated patches are removed. All distributions have a peak close to 200 A.U. }
  \label{fig:histograms_valence_I}
\end{figure}

\begin{figure}[h]
  \centering
  \includegraphics[width=12cm]{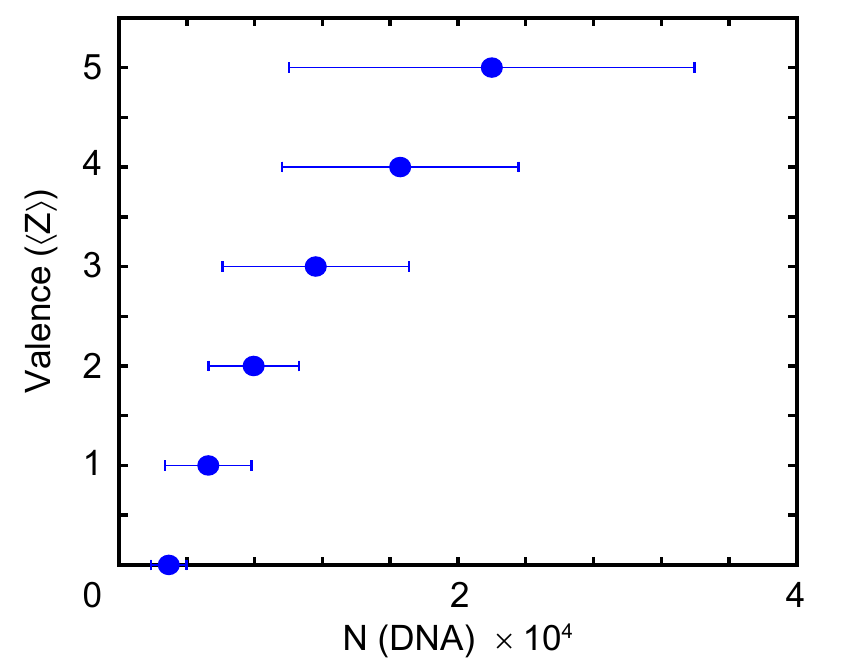} % requires the graphicx package
  \caption{A plot of the estimated total DNA on the droplets, $N$ for droplets with a given valence $Z$ for all droplets after droplets with supersaturated patches are removed from consideration.}
  \label{fig:valence_I}
\end{figure}

Upon examining the data, we noticed that the distribution of intensities for adhesion patches has a long tail (see Fig.~\ref{fig:histograms_I}), with a small fraction of patches showing extraordinarily high intensities.
We instituted a threshold of the median patch intensity plus twice the standard deviation of the intensity distribution (approximately 2100 A.U.), exceeded by ~20\% of patches.
All droplets containing a patch above this threshold were removed from consideration, under the hypothesis that they contained supersaturated patches that were kinetically frustrated from seeding further patches.
Upon removal, the new distributions for intensity still show a peak near 200 A.U. for all valences (see Fig.~\ref{fig:histograms_valence_I}).
The resulting plot of the valence, $Z$, with the estimated amount of DNA per droplet, $N$, after this cut off is applied shows that $Z$ now increases much more linearly with $N$ (see Fig.~\ref{fig:valence_I}).
These results are insensitive to the exact threshold value, as long as the excluded patches have intensities well above the median.

